\documentclass{eptcs}

\AtBeginDocument{%
  \providecommand\BibTeX{{%
    \normalfont B\kern-0.5em{\scshape i\kern-0.25em b}\kern-0.8em\TeX}}}

\usepackage[inference]{semantic}
\usepackage{breakurl}
\usepackage{amsmath}
\usepackage{amssymb}
\usepackage{url}

\usepackage{caption}
\usepackage{graphicx}
\usepackage{subcaption}
\usepackage{listings}
\newcommand{\sC}[1]{\textrm{\textup{\fontfamily{cmss}\selectfont \textbf{#1}}}}
\newcommand{\sco}[1]{\textrm{\textup{\fontfamily{cmss}\selectfont \texttt{#1}}}}

\newcommand{\ltrue}{\textrm{\(t\kern-0.1em t\)}}

\newcommand{\namedRule}[2]{
	\textrm{\footnotesize [#1]: \ \ \ } &
	#2
}

\title{A Tool for Describing and Checking Natural Semantics Definitions of Programming Languages}

\author{Georgian-Vlad Saioc
\institute{Department of Computer Science\\ Aarhus University, Denmark}
\email{gvsaioc@cs.au.dk}
\and
Hans Hüttel
\institute{Department of Computer Science \\ Aalborg University, Denmark\\[2mm]
Department of Computer Science \\  University of Copenhagen, Denmark}
\email{hans.huttel@di.ku.dk}
}
\def\titlerunning{A Tool for Describing and Checking Natural Semantics Definitions}
\def\authorrunning{G.-V. Saioc and H. Hüttel}

\begin{document}
\maketitle 

\begin{abstract}
Many universities have courses and projects revolving around compiler or interpreter implementation as part of their degree programmes in computer science. In such teaching activities, tool support can be highly beneficial. While there are already several tools for assisting with development of the front end of compilers, tool support tapers off towards the back end, or requires more background experience than is expected of undergraduate students.

Structural operational semantics is a useful and mathematically simple formalism for specifying the behaviour of programs and a specification lends itself well to implementation; in particular big-step or natural semantics is often a useful and simple approach. However, many students struggle with learning the notation and often come up with ill-defined and meaningless attempts at defining a structural operational semantics. A survey shows that students working on programming language projects feel that tool support is lacking and would be useful.

Many of these problems encountered when developing a semantic definition are similar to problems encountered in programming, in particular ones that are essentially the result of type errors.

We present a pedagogical metalanguage based on natural semantics, and its implementation, as an attempt to marry two notions: a syntax similar to textbook notation for natural semantics on the one hand, and automatic verification of some correctness properties on the other by means of a strong type discipline.
The metalanguage and the tool provide the facilities for writing and executing specifications as a form of programming. The user can check that the specification is not meaningless as well as execute programs, if the specification makes sense. 

\end{abstract}

\section{Introduction}

Many degree programmes in computer science have a substantial activity devoted to programming language design and implementation, both of which may be aided by structural operational semantics. As a motivating example, both undergraduate degree programmes in computing at Aalborg University involve such a project in the $4^{\textrm{th}}$ semester, where the learning goals involve relating the definition and implementation.

Providing tool support for students carrying out such projects is an important challenge. There are several tools for parser generation such as ANTLR \cite{Parr13}, Happy\cite{Happy}, Menhir\cite{menhir}, or Yacc \cite{lexyacc}, that support the definition of programming language syntax, with students using them extensively, to the point of becoming a central part of their project work. However, experience shows that many students struggle with writing a meaningful definition of a structural operational semantics, where some students even attempt to use parser generators as a surrogate for enforcing semantics. In addition, project supervisors often come from backgrounds outside the area of programming languages and are consequently often unfamiliar with the topic. 

Moreover, tools that use operational semantics are few and far between. Early attempts include RML \cite{RML} and Letos \cite{letos}. The most ambitious of its kind is PLT Redex\cite{plt-redex} which uses the approach to reduction semantics advocated by Felleisen et al. In PLT Redex, specifications are given in the Racket language. The advantage of this approach is clear: writing a definition of the semantics of a programming language becomes similar to programming in a Lisp-like language. The disadvantage is that learners of formal semantics have to balance both the mathematical notation and the syntax of the Racket language -- and must learn both. The K Framework \cite{DBLP:journals/iandc/SerbanutaRM09} and its implementation in Maude \cite{VERDEJO2006226} give an alternative computational model for operational semantic definitions. While the K Framework is general in nature, it is not yet widely used in teaching, and the focus of this framework is not that of structural operational semantics, but a new setting based on Rewriting Logic; its major concerns differ from that of our work, instead focusing on modularity within operational semantics.

Natural semantics is a useful entry point for learning about the formal semantics of programming languages, and in the present work we focus on this style of semantics. In our experience many of the commonly encountered obstacles stem from misunderstandings that are really programming errors in the setting of the notation used. In particular, many such misunderstandings can best be viewed as type errors.

In this paper we describe a tool in the form of a metalanguage based on natural semantics\cite{Khan1988}. The syntax of the metalanguage attempts to reflect usual textbook notation, better aligning a learner's curriculum and tool usage. This is supplemented by a type system that allows learners to quickly discover inconsistencies in proposed definitions. To test its capabilities, the metalanguage has been used to implement definitions in an existing textbook 
\cite{DBLP:books/daglib/0040165}, uncovering several inconsistencies.

In Section \ref{sec:sos} we outline the principles underlying structural operational semantics and use this account to describe typical challenges in Section \ref{sec:challenges}. This is followed by a metalanguage description in Section \ref{sec:metalanguage},
examples of how the tool has been used to discover flaws in structural operational semantics definitions from a textbook in Section \ref{sec:textbook-errors},
and details on the implementation in Section \ref{sec:implementation}. Conclusions and future work are discussed in Section \ref{sec:future-work}.

\section{A survey of students' experiences}

We carried out a survey among students that had recently completed their undergraduate degrees in computer science at Aalborg University to find out more about their experience with the programming language project in the 4th semester. As previously noted, usage of structural operational semantics was commonplace in the projects.

We asked four questions concerning the importance of tool support in this setting. Each of which could be answered according to a 5-point Likert scale. 

There were a total of 25 respondents. An overview of the reponses is shown in Figure \ref{fig:response}.

\begin{figure*}
    \centering
    \begin{tabular}{|p{4cm}|p{1.4cm}|p{1.4cm}|p{1.4cm}|p{1.4cm}|p{1.3cm}|}
    \hline & Strongly agree & Somewhat agree & Neither & Somewhat disagree & Strongly disagree \\
    \hline 
    The facilities for checking if the semantic definition of the programming language of the project were satisfactory. & 13 & 17 & 29 & 38 & 4 \\
\hline  It was easy to determine exactly what the problems in our attempt at a semantic definition were. & 8 & 4 & 38 & 38 & 13 \\
\hline The tool support for devising a semantic definition was satisfactory. & 4 & 13 & 21 & 29 & 33 \\
\hline It would be useful if one could generate a prototype interpreter based on a semantic definition. & 42 & 50 & 4 & 4 & 0\\
\hline
\end{tabular}
    \caption{Results of the survey. Numbers represent the percentage of respondents}
    \label{fig:response}
\end{figure*}

The numbers indicate that students feel that tool support for building and checking definitions in structural operational semantics was lacking and would have been desirable in a project setting.

We also asked for qualitative comments. The replies that we got support this impression. The general observation was that a tool that can check for consistency of a semantic definition and allow one to get an overview of it is important. Some students also mentioned that it would be useful to be able to typeset transition rules in \LaTeX\ in a more structured fashion.

Below are two sample quotes (translated from Danish) that highlight the major issues:
\begin{itemize}
    \item 
\begin{quote}
   Our greatest challenges were those of checking if all the semantic rules were correctly constructed (consistent) and of making the choice of which parts of the semantics would be responsible for what.
\end{quote}
    \item
\begin{quote}
    When I followed the course back in 2018, there were no tools that we used. You had to go through the semantics yourself to spot mistakes or go through the semantics together with someone else. This walkthrough was indeed helpful (until the rules became too long) but one could easily end up staring blindly at the rules.
\end{quote}
\end{itemize}
\section{Structural Operational Semantics} \label{sec:sos}

Structural operational semantics (SOS) is a formalism for describing programming language semantics, others being denotational semantics\cite{densem} and attribute grammars. The approach originates with the work by Plotkin \cite{plotkin} and differs from the others in that a semantic definition describes the computation steps of a program by means of a syntax-directed definition using inference rules. SOS comes in two variants: small-step and big-step (natural) semantics. What follows is an overview of SOS, with the focus being on natural semantics, its components, and the usual textbook notation, which comes into play when describing the metalanguage.

\subsection{Overview}

SOS revolves around defining \emph{transition systems}, relating execution states of a program, also known as \textit{configurations}. A transition typically involves the syntactic element to be evaluated, and the current bindings. In a small-step semantics, a transition represents a single evaluation step, with successive iterations leading to a final evaluation, whereas in a natural semantics, transitions represent complete evaluations.

The formal definition of a transition system in a natural semantics is as a triple, $(I, \to, F)$, where $\to$ expresses a relation between initial and final configurations, represented by sets $I$ and $F$, respectively. Intuitively, for some $i \in I$ and $f \in F$, if transition $i \to f$ holds, then $i$ evaluates to $f$. The relation between configurations is inferred via a \emph{transition system}, the rules of which are defined according to the following template:
\begin{align}
\label{rule-template}
    \namedRule{RULE NAME}{
        \inference{
            i_1 \to f_1  & ... &
            i_n \to f_n
        }{
            i \to f
        }
    } & \ \ \emph{optional side conditions}
\end{align}
The \textit{conclusion}, denoted by judgement $i \to f$, holds if all premises, $i_i \to f_i$, for $1 \leq i \leq n$, hold. An inference rule with no premises is an \emph{axiom}, and holds as long as all side conditions hold. A rule may have side conditions, typically describing predicates on configurations in the rule. By generalisation, any $i \to f$ holds if a \emph{closed derivation tree} may be inferred through the transition rules. For any derivation tree we have that:
\begin{itemize}
    \item The tree is rooted at the conclusion of a rule,
    \item Successfully applying a rule generates a sub-tree for each premise e.g., each $i_i \to f_i$ in (\ref{rule-template}) may produce a corresponding sub-tree,
    \item Satisfied axioms are leaves in the tree.
\end{itemize}
A derivation tree is defined inductively as closed if it is a satisfied axiom, or all of its sub-trees are closed. If a closed derivation tree is produced for some transition, $i \to f$, then the transition satisfies the semantics of the language.

Transitions may involve an \textit{antecedent}, denoted as $\alpha \vdash i \to f$, read as \textit{given $\alpha$, we have that $i \to f$}.

An SOS is syntax-directed, where transition rules are (typically) exhaustively defined over all syntactical members of the language. This entwines the semantics of the language with its abstract syntax, making the formation rules of the abstract grammar an important component of a semantics definition.

\subsection{An example specification}
\label{lang-exmp}

Consider a small language, \textbf{Imp}, in conventional textbook notation, consisting of sequences of arithmetic expressions and variable declarations. The syntactic categories are \sC{Stm}, \sC{Exp}, \sC{Var} and \sC{Num} for statements, expressions, variable names and integer literals, respectively. The grammar is presented in Figure \ref{fig:small-language-semantics} (a), where $S \in \sC{Stm}$, $e \in \sC{Exp}$, $x \in \sC{Var}$, and $n \in \sC{Num}$.
A natural semantics that matches the informal description of this language can be expressed through transition systems for expressions and statements, respectively, given in Figures \ref{fig:small-language-semantics} (b) and (c).

\begin{figure}[h]
    \centering
    \begin{subfigure}[b]{\textwidth}
\begin{align*}
    S & ::=\ S_1\ \sco{;}\ S_2\ |\ x\ \sco{=}\ e\\
    e & ::=\ n\ |\ x\ |\ e_1\ \sco{+}\ e_2\ 
\end{align*}
\caption{Grammar}
\label{fig:imp-grammar}
    \end{subfigure}
    \begin{subfigure}[b]{\textwidth}
    \begin{align*}
    \namedRule{CONST}{
        s \vdash n \to_e v
    } & \textrm{where } n \textrm{ denotes } v\\\\
    \namedRule{VAR}{
        s \vdash x \to_e v
    } & \textrm{where } s(x) = v\\\\
    \namedRule{ADD}{
        \inference{
            s \vdash e_1 \to_e v_1 &
            s \vdash e_2 \to_e v_2
        }{
            s \vdash e_1\ \sco{+}\ e_2 \to_e v
        }
    } & \textrm{where } v = v_1 + v_2
    \end{align*}
    \caption{Transition rules for expressions}
    \end{subfigure}
    \begin{subfigure}[b]{\textwidth}
    \begin{align*}
    \namedRule{DECL}{
        \langle x\ \sco{=}\ e,\ s \rangle \to_S s[x \mapsto v]
    } & \textrm{where } s \vdash e \to_e v
    \\\\
    \namedRule{COMP}{
        \inference{
            \langle S_1, s \rangle \to_S s' &
            \langle S_2, s' \rangle \to_S s''
        }{
            \langle S_1 \sco{;}\ S_2, s \rangle \to_S s''
        }
    }
    \end{align*}
    \caption{Transition rules for statements}
    \end{subfigure}
    \caption{Subset of the syntax and semantics of \textbf{Imp}}
    \label{fig:small-language-semantics}
\end{figure}

We define the transition system of expressions as $(\sC{Exp}, \to_e, \mathbb{Z})$. An example axiom is VAR, denoting variable lookup, where metavariable $s$ is the current environment, binding variables to values. We define $s \in$ \textsf{E}, where $\textsf{E}$ is $\sC{Var} \rightharpoonup \mathbb{Z}$. Axiom VAR can be applied for all $s$ and $x$ as long as $s(x)$ is defined. Evaluating a binary operation requires a rule instead, as it depends on the results of the operands obtained from evaluating each sub-expression. One such example is the ADD rule, where the sum of the results $v_1$ and $v_2$, produced by the premises (the respective evaluations of sub-expressions $e_1$ and $e_2$), is the result of evaluating an addition operation in \textbf{Imp}.

The transition system for statements, $(\sC{Stm}\ \times\ \textsf{E}, \to_S, \textsf{E})$, involves updating variables in the environment. The axiom is DECL, showing how a declaration leads to updating $x$ to $v$ in environment $s$. Compared to the transition system of expressions, $s$ is now part of the configuration, as opposed to the antecedent. The transition system of expressions is referenced by using $\to_e$ in the side conditions of DECL and EXP. The only rule, COMP, expresses statement composition. It shows how the environment is updated between statements $S_1$ and $S_2$, where $s'$, produced by $S_1$ is used in the initial configuration of $S_2$. 

\section{Challenges with mastering the notation} \label{sec:challenges}

While the mathematical machinery underlying structural operational semantics is fairly modest, learning the notation and using it correctly can be challenging for students. The following reflections arise from our teaching experience.

By far the most common problem is not realising that all transitions within the rules of a transition system must follow the same format. This appears to be what Meyer and Land call a \emph{threshold concept} \cite{Meyer03thresholdconcepts} -- a central concept that transforms how one perceives a given subject once understood, but might initially present a steep learning curve.

Consider the declaration of functions in an imperative programming language with static scope rules. The binding model for variables uses a mapping $env_V$, binding variables to addresses, and a store $sto$, binding addresses to values. The binding model for functions is a function environment $env_F$, a finite mapping that binds function names $f$ to triples $\langle S, env_V,env_F\rangle$. If $\sC{FNames}$, $\sC{Stm}$, $\textsf{EnvV}$ and $\textsf{EnvF}$ denote the types of function names, statements, variable environments and function environments, respectively, then we define all $env_F \in \textsf{EnvF}$ as functions binding function names in the language to closures at run-time. All $env_F$ have the following signature:
\[ env_F: \sC{FNames} \rightharpoonup \sC{Stm} \times \textsf{EnvV} \times \textsf{EnvF} \]
For any closure, we have that $S$ is the body of the function and $env_V \in \textsf{EnvV}$ and $env_F$ are the variable and function environments captured during the definition of $f$, binding its free-variables. 

The semantics is modelled as a transition system, where function declarations, denoted by $D_F$, return an updated function environment. The grammar of $D_F$ is:
\[ D_F\ =\ \sco{fun}\ f\ \sco{is}\ S \sco{;}\ D_F\ |\ \epsilon \]
and transitions over $D_F$ should follow the format:
\[ env_V \vdash \langle D_F, env_F \rangle \to env'_F \]

Figure \ref{fig:incorrect-rule-example} showcases both the correct rules (a), and the erroneous definition given by one student (b). The latter highlights many common mistakes made when using the notation:

\begin{figure}
    \centering
    \begin{subfigure}[b]{\textwidth}
\begin{align*}
    \namedRule{DECL}{
        \inference{
            env_V \vdash \langle D_F,env_F[f \mapsto \langle S,env_V,env_F \rangle] \rangle \to env'_F
        }{
            env_V \vdash \langle \texttt{fun}\; f \; \texttt{is}\; S \sco{;}\ D_F, env_F \rangle \to env'_F
        }
    }\\\\
    \namedRule{EMPTY}{
        env_V \vdash \langle \epsilon, env_F \rangle \to env_F
    }
\end{align*}
    \caption{Correct function definition rules}
    \end{subfigure}
    \hfill
    \begin{subfigure}[b]{\textwidth}
\begin{align*}
    \namedRule{DECL}{
        \inference{
            \langle D_F,env_F[f \mapsto S] \rangle \to env'_F
        }{
            env_V \vdash \langle \texttt{fun}\; f \; \texttt{is}\; S, sto, env_F \rangle \to \langle sto, env'_F \rangle
        }
    }
\end{align*}
    \caption{Incorrect, student-defined rule}
    \end{subfigure}
    \caption{Operational semantics of function definitions in an imperative language with static scoping rules}
    \label{fig:incorrect-rule-example}
\end{figure}
\begin{enumerate}
\item \label{issue1} The formation rules of $D_F$ are not properly represented in the initial configuration of the conclusion. While the individual function declaration is present, the trailing $D_F$ is not.
\item \label{issue2} In the premise, the function environment is updated as  $env_F[f \mapsto S]$, where the function is only bound to its body while omitting the function and variable environments required by the closure.
\item \label{issue3} The transitions in the premise and the conclusion have different formats. The transition in the premise is not relative to a variable environment, unlike the one in the conclusion.
\item \label{issue4} The final configurations differ. For the premise, the final configuration is a function environment $env'_F$ , whereas in the conclusion, the terminal configuration is a pair $(sto,env'_F)$. The purpose of including the store at all is not entirely clear.
\end{enumerate}

All of these issues are often encountered in ill-defined transition rules proposed by students, and have counterparts in programming. Fortunately, all of these errors can be captured statically:
\begin{itemize}
    \item All issues can be thought of in part as \textit{type errors}, e. g., the omission of $D_F$ in issue \ref{issue1} as a malformed constructor call, or transitions and configurations with different formats.
    \item Due to the omission of $D_F$, highlighted in issue \ref{issue1}, its usage is also an inadvertent instance of \textit{using an undeclared variable}.
\end{itemize}




Since the students that make these kinds of mistakes are in their second year of an undergraduate degree programme in computing, they already have experience in dealing with such issues in the setting of programming. This suggests they would benefit from a dedicated programming language with a static semantics (including a type system) capable of capturing these kinds of errors and displaying them in a more familiar format.

\section{Metalanguage}
\label{sec:metalanguage}


The metalanguage was inspired by one of the key notions used by most lexer and parser generators, specifically the usage of a domain-specific language. In the case of lexer and parser generators, their respective domain-specific languages codify regular expressions and context-free grammars, using the corresponding pen-and-paper formalism as a basis. The metalanguage follows this idea, but opts for modelling language semantics, instead, using structural operational semantics, namely the big-step variant, as a basis. This choice of formalism follows naturally, given the stated pedagogical goals and aforementioned utility of structural operational semantics when used by students for language design. As such, it is tailored towards expressing the essential elements found in natural semantics specifications of languages, namely transition systems and abstract syntax definitions.

Syntactically, the metalanguage is modelled for \textit{writing specifications close to textbook notation}. Since SOS notation is better accommodated by rich text editors, some compromises are made to keep a balance between visual fidelity and writability.
Specifications are {checked for type safety}, achieved through a strong typing discipline (sections \ref{metalanguage-domains} and \ref{metalanguage-syntax}). Students may also specify concrete expressions to evaluate, for which to inspect the results.

A specification in the metalanguage is a sequence of definitions, optionally followed by evaluations. The following subsections describe each type of definition, along with related concepts and examples based on \textbf{Imp}, defined in section \ref{lang-exmp}.

\subsection{Domains}
\label{metalanguage-domains}

The metalanguage provides a monomorphic type system, where types are represented as domains. Basic domains are one of \textbf{Int}, \textbf{String}, \textbf{Bool}, or \textbf{Symbol}, where the former three are typical of other languages, while the latter is specifically introduced to represent variables in the specified language. String values are enclosed in double quotes, while symbols are enclosed in back-ticks. In practice, symbols are strings defined over an alphabet idiomatic to variable naming (alphanumerics, underscore, etc.).

Domain composition is achieved similarly to strongly typed functional languages\cite{Haskell,OCaml}. Composed domains can be one of product, arrow, or disjoint unions, the latter of which can be used to declare algebraic data types. An example would be defining a domain for lists of integers:
\begin{verbatim}
domain List = { nil + cons : Int * List };
\end{verbatim}
Users can use the two constructors \texttt{nil} and \texttt{cons} to construct lists inductively.

Product and arrow types can be aliased, or inlined whenever type annotations are required, but disallow inductive definitions, whereas disjoint unions cannot be aliased or inlined, but may be inductive, like the \texttt{List} example.

The metalanguage interpreter has limited type inference capabilities, requiring type annotations only on some expressions, such as formal parameters in functions, and headers of transition systems. In \textbf{Imp}, the domain of environments can be defined as an arrow type from symbols to integers:
\begin{verbatim}
domain Env = Symbol -> Int;
\end{verbatim}

\subsection{Syntax}
\label{metalanguage-syntax}

Syntax definitions are functionally identical to union domains, but use special syntax, by combining domain definitions with symbols, enclosed in single quotes and denoting purely syntactical elements (e.g. keywords). This allows expressing abstract syntax in a way that approximates textbook grammar definitions using BNF notation. A syntax constructor is derived based on the sequence of symbols and domain placements, where domains are replaced by \textit{holes} (denoted as \_). Syntax constructors in a specification must be unique in a specification, and may omit symbols entirely as long as no other production rule with no symbols has the same number of domains. The metalanguage interpreter adapts its output based on this distinction (e. g. for reporting errors or pretty printing).

The syntax of \textbf{Imp} can be defined as:
\begin{lstlisting}
syntax Stm = Symbol '=' Exp
    | Stm ';' Stm;
syntax Exp = '#' Int
    | Symbol
    | Exp '+' Exp;
\end{lstlisting}
Note that for \texttt{Exp}, integer constants are defined as an integer value preceded by \texttt{'\#'}, whereas variables are denoted as simply a symbol value without any additional decoration. Binary addition expressions consist of two sub-expressions and the \texttt{'+'} symbol in infix notation.

\subsection{Data and expressions}
\label{metalanguage-data}

Data can be defined similarly to most functional languages by binding variables to expressions, given as an extended simply-typed $\lambda$-calculus. Expression are also used at various points in the specification, such as side conditions in transition systems, making specifications a natural semantics interpolated with $\lambda$- calculus.

Extension to the $\lambda$-calculus include function updates and the bottom element. For \textbf{Imp}, one could define an empty environment as:
\begin{verbatim}
let empty = \x : Symbol . -|In|;
\end{verbatim}
This definition makes \texttt{empty} a mapping from symbols to integers. The body of the $\lambda$-expression is the bottom element, annotated with type \textbf{Int}, so as to determine its expected type. The bottom element allows users to define partial functions, and generates an exception when evaluated at run-time. One can express binding updates throughout a specification (e. g. in rules). An example of an update expression would be \texttt{empty[`x` -> 0]}, evaluating to a closure which yields 0 when applied to symbol \texttt{`x`} while the other bindings are unaffected. Currently, only functions with formal parameters inhabiting basic domains can be updated.

Syntax expressions are defined as values enclosed between braces, (\texttt{\{}, \texttt{\}}). For a syntax expression to be valid, it must be a sequence of syntactical symbols and metavariable placements that matches one of the defined syntax constructors: e.g., \texttt{\{e1 '+' e2\}} is a syntax expression denoting addition for \textbf{Imp}, where \texttt{e1} and \texttt{e2} are metavariables representing the sub-expressions.

\subsection{Transition systems}

The final definition construct involves transition systems. We can now define the semantics of \textbf{Imp}, starting with the transition system for expressions, named \texttt{e}. An example subset of the transition system may be given as:
\begin{verbatim}
system e : Env |- Exp ==> Int =
...
    [[ CONST ]]: s |- {'#' n} ==> n;
    
    [[ VAR ]]: s |- {x} ==> s(x);

    [[ ADD ]]: s |- {e1 '+' e2} ==> v1 + v2 \\
        s |- e1 ==> v1,
        s |- e2 ==> v2;
...
end
\end{verbatim}
The definition is prefixed by a specially formatted type annotation, expressing the domains of the antecedent (optional), initial and final configurations. Transition system \texttt{e} accepts members of \texttt{Exp} as its initial configurations, and evaluates to integers. Since the antecedent type is \texttt{Env}, only functions from symbols to integers are valid antecedents.

The components of a transition rule are:
\begin{enumerate}
    \item \textit{The label}, in double brackets, which names a rule. Labels are purely cosmetic, and help with pretty-printing.
    
    \item \textit{The conclusion}, represented by the first transition in a rule. Conclusions have patterns on the left-hand side of the arrow, for the antecedent and initial configuration, and an expression on the right-hand side. Patterns may be formed from variables, constants, pairs, and union/syntax constructors. When applying a rule, pattern matching is attempted first, followed by evaluating the premises, and concluding with evaluating the final configuration.
    
    \item \textit{The premises}, if present, are separated from the conclusion through \texttt{\textbackslash\textbackslash}, and from each other through commas. For homogeneity, transitions, side conditions, and local declarations are all considered premises, and are interpreted in order of declaration. Premise transitions flip the occurrence of patterns and expressions, where the left-hand side is given as an expression to be evaluated, and the final configuration is a pattern that the resulting value must match. 
\end{enumerate}

The transition system for statements would be given as:
\begin{verbatim}
system S : Stm * Env ==> Env =
    [[ DECL ]]: ({x '=' e}, s) ==> s[x -> v] \\
        s |- e =e=> v;
    
    [[ COMP ]]: ({S1 ';' S2}, s) ==> s'' \\
        (S1, s) ==> s',
        (S2, s') ==> s'';
end
\end{verbatim}
As in the definition proper, the antecedent is absent, and configurations are given as pairs between statement syntax and states. If premise transitions refer to the same system they are defined in, it suffices to use \texttt{==>}. If referring to another transition system, the notation used is \texttt{=}\emph{System name}\texttt{=>}. The DECL rule for \textbf{Imp} statements references transition system \texttt{e} to denote the evaluation of the right-hand side of the variable declaration. 

\subsection{Evaluations}

Following definitions, a specification may include evaluations. The metalanguage interpreter evaluates expressions by simulating interpretations using definitions in the specification, and print their results. Consider the following transition:
\begin{align*}
[x \mapsto 5] \vdash 3\ \sco{+}\ x \to_e 8
\end{align*}
An evaluation for it would be written as
\begin{verbatim}
evaluate empty[`x` -> 5] |- {{'#' 3} '+' {`x`}} in e
\end{verbatim}
Since the root of the syntax tree is a (\texttt{\_ '+' \_}) node, an application of the ADD rule is attempted. This is followed by successive applications of the CONST and VAR rules for each respective child. The result is the expected value, 8.

\section{Flawed definitions in textbooks}
\label{sec:textbook-errors}

Pen-and-paper SOS specifications get increasingly difficult to keep
consistent the more complex they become, with subtle errors sneaking
into otherwise sound specifications, even for experienced users. Below
are some examples of errors in the specifications languages from a textbook on SOS \cite{DBLP:books/daglib/0040165}, and how the
metalanguage interpreter captured them:

\begin{itemize}
\item For \textbf{Bur}, an imperative language
with records and procedure calls, the side-conditions of WHILE rules contain ill-formed
  transitions for evaluating guards. Since \textbf{Bur} contains
  records, variables and procedures, there is a corresponding
  environment for each. The former two are required in the antecedent
  for boolean expression evaluations, but the side conditions omit the
  record environment:
     \begin{align*}
         & \textrm{if } e_V, sto \vdash b \to_b true.
     \end{align*}
     Transcribing the specification of \textbf{Bur} will result in the side conditions emitting a type error, where the type of boolean expression transitions is defined as:
     \begin{verbatim}
system b : EnvR * EnvV * Sto |- Bexp ==> Bool\end{verbatim}
     \item The semantics of block statements in \textbf{Bur} is defined as:
     \begin{align*}
         \namedRule{BLOCK}{
             \inference{
                 e_R \vdash \langle D_V, e_V, sto \rangle \to_{DV} (e_V', sto'') \\
                 e_V' \vdash \langle D_R, e_R \rangle \to_{DR}(e_R', e_V'') \\
                 e_V'' \vdash \langle D_P, e_P \rangle \to_{DP} e_P' \\
                 e_R', e_V'', e_P' \vdash \langle S, sto'' \rangle \to sto'
             }{
                 \begin{array}{ll}
                      e_R, e_V,
                      e_P  \\
                 \end{array} \vdash \langle \sco{begin}\ D_V\ D_P\ D_R\ S\ \sco{end}, sto \rangle \to sto'
             }
}
            \end{align*}
     consisting of variable, procedure and record declarations, followed by statement execution. The rule breaks the previously established format for transitions in $\to_{DR}$, for record declarations. The type of transitions in $\to_{DR}$ would be defined as:
 \begin{verbatim}
system dR : DR * EnvR * EnvV ==> EnvR * EnvV\end{verbatim}
     but the block rule shifts the variable environment to the antecedent, as opposed to the initial configuration. The procedure declaration premise, denoted by $\to_{DP}$, has a similar issue, where the type would be defined as:
\begin{verbatim}
system dP : EnvV * EnvR |- DP * EnvP ==> EnvP\end{verbatim}
    but, in the above rule, the record environment, denoted by \texttt{EnvR}, is missing in the antecedent. Both discrepancies result in type errors.
     \item In the language \textbf{Flan}\cite{DBLP:books/daglib/0040165}, a simple functional language modelled after the $\lambda$-calculus, values used as final configurations in the semantics definition, are either constants, closures or recursive closure. The formal definition for this set is given as:
     \begin{align*}
         \sC{Values} = &\ \sC{Fcon} \cup (\sC{Var} \times \sC{Fexp} \times \sC{Env}) \cup (\sC{Var} \times \sC{Var} \times \sC{Fexp} \times \sC{Env})
     \end{align*}
     This set may expressed as a grammar, with each member of the union as a formation rule\footnote{Syntax symbols may use UTF-8 code points.}:
 \begin{lstlisting}
syntax Values = Fcon
| '$\lambda$' Symbol '.' Fexp ',' Env
| '$\lambda$' Symbol Symbol '.' Fexp ',' Env;
 \end{lstlisting}
     Constants are simply represented as set \texttt{FCon}, while closures are $\lambda$ terms with symbols denoting the formal parameters (one symbol for non-recursive closures, and two symbols for recursive closures) and expressions as bodies, and the captured environment.
    The type of transitions of \textbf{Flan} could be given as:
 \begin{lstlisting}
system Flan : Env |- Exp ==> Values
 \end{lstlisting}
     However, the textbook rules of \textbf{Flan} also include the following rule for pairs:
 \begin{align*}
     \namedRule{PAIR}{
         \inference{
             env \vdash e_1 \to v_1 &
            env \vdash e_2 \to v_2
         }{
            env \vdash (e_1, e_2) \to (v_1, v_2)
         }
     }
 \end{align*}
     Pairs are not covered under \sC{Values}. When attempting to transcribe the rule for pairs, this issue will become apparent, and it can be easily fixed by extending the set of values with the formation rule \texttt{'(' Values ',' Values ')'}.
 \end{itemize}

\section{Implementation}
\label{sec:implementation}

The metalanguage interpreter was implemented in Haskell\cite{Haskell}. As an executable it can either be used as a rudimentary REPL, or by targetting a specification in a separate file. 

\subsection{Checking the specification}

At run-time, the tool performs several traversals over the specification syntax tree. In order, the traversals involve:

\begin{itemize}
    \item \textit{Domain analysis}, such as checking for free domain variables and that domains are well-formed.
    \item \textit{Type analysis}, such as type checking expressions and transition systems, and inferring types.
    \item \textit{Evaluation} of all data definitions in order of declaration, followed by simulating interpretations for each \texttt{evaluate} construct. Section \ref{simulate-interp} covers the evaluation strategy.
    \item \textit{Code generation}, targeting \LaTeX\ for the back-end. Compared to a dedicated tool like LETOS\cite{letos}, the generated \LaTeX\ is not publication quality, but is reasonable enough to expedite a report writing process for students. A subset of the generated code for \textbf{Imp} expressions is given in Figure \ref{fig:latex-output}\footnote{Transitions in generated code are denoted as $\Downarrow$ as opposed to $\to$. The set $\chi^{*}$ is the domain of symbols.}.
    
\end{itemize}
\begin{figure}
    \centering
\begin{align*}
Exp & ::= \#\ \mathbb Z\\
&|\ \chi^{*}\\
&|\ Exp\ \sco{+}\ Exp\\\\
\namedRule{VAR}{
s \vdash x \Downarrow s(x)
 & }\\\\
\namedRule{CONST}{
s \vdash \sco{\#}\ n  \Downarrow n
 & }\\\\
\namedRule{ADD}{
\inference{
s \vdash e_{1} \Downarrow v_{1}\\
s \vdash e_{2} \Downarrow v_{2}\\
}{
s \vdash e_{1}\ \sco{+}\ e_{2} \Downarrow v_{1} + v_{2}
} & }
\end{align*}
    \caption{Example \LaTeX{} output for the expression grammar and transition rules}
    \label{fig:latex-output}
\end{figure}

\subsection{Executing a specification}
\label{simulate-interp}

Evaluations are carried out by attempting to build a derivation for a given transition. This is achieved by successively applying the rules in a transition system until one that matches the pattern is found. This is followed by recursively evaluating each premise, in the order of declaration. If no rules are successfully applied, e. g. no pattern was matched successfully or no rules satisfy all premises, the interpretation is abandoned, displaying an error message tracing all attempted rule applications.

Multiple rules may have the same initial configuration pattern in the conclusion. Consider extending the grammar of \textbf{Imp} with conditional statements as in Figure \ref{fig:imp_cond}.

\begin{figure}
    \centering%
\begin{subfigure}[b]{\textwidth}
\begin{align*}
    S & ::=\ \cdots\ |\ \sco{if}\ e\ \sco{then}\ S_1\ \sco{else}\ S_2
\end{align*}
\begin{align*}
\namedRule{IF-TRUE}{
    \inference{
        \langle S_1,\ s \rangle \to_S s'
    }{
        \langle \sco{if}\ e\ \sco{then}\ S_1\ \sco{else}\ S_2, s \rangle  \to_S s'
    }
} & \textrm{if } e \to_e v,  \textrm{ where } v \neq 0\\\\
\namedRule{IF-FALSE}{
    \inference{
        \langle S_2,\ s \rangle \to_S s'
    }{
        \langle \sco{if}\ e\ \sco{then}\ S_1\ \sco{else}\ S_2, s \rangle \to_S s'
    }
} & \textrm{if } e \to_e 0
\end{align*}
    \caption{Conditional statements syntax and natural semantics}
\end{subfigure}

\begin{subfigure}[b]{\textwidth}
\begin{lstlisting}
syntax Stm = ... | 'if' Exp 'then' Stm 'else' Stm;

system S : Stm * Env ==> Env =
    ...
    [[ IF-TRUE ]]: ({'if' b 'then' S1 'else' S2}, s) ==> s' \\
        b =e=> v,
        if v != 0,
        (S1, s) ==> s';
    
    [[ IF-FALSE ]]: ({'if' b 'then' S1 'else' S2}, s) ==> s' \\
        b =e=> 0,
        (S2, s) ==> s';
    ...
\end{lstlisting}
    \caption{Metalanguage definition for conditional statements in \textbf{Imp}}
\end{subfigure}
    \caption{Extending \textbf{Imp} syntax and semantics with conditional statements}
    \label{fig:imp_cond}
\end{figure}

At run-time, the tool applies \texttt{IF-TRUE} first. If the initial match succeeds, this is followed by evaluating \texttt{b =e=> v}. If the result is not 0, the interpreter proceeds with evaluating \texttt{S1}, concluding with \texttt{s'} if successful. If \texttt{v} is 0, the application of \texttt{IF-TRUE} is abandoned, and the interpreter proceeds to look for the next matching rule, namely \texttt{IF-FALSE}.

Rules need not be exhaustive: some syntactic constructs may not have transition rules at all or some side conditions may not be covered. This can be a deliberate decision on the language designer's part, an example being:
\[ S :: = \cdots\ |\ \texttt{abort} \]
where $\texttt{abort}$ is a construct with no defined rule that causes the evaluation to terminate prematurely.

\subsection{Limitations}

Due to the chosen evaluation strategy, declaration order matters for rules. Take as an example the rule CIRCULAR in Figure \ref{fig:circular-rules}.
\begin{figure}
\begin{align*}
    e ::= ...\ |\ \sco{stop}
\end{align*}
\begin{align*}
\namedRule{CIRCULAR}{
    \inference{
        e \to v
    }{
        e \to v
    }
}\\
    \namedRule{STOP}{
        \sco{stop} \to 0
    }
\end{align*}
    \caption{Example of non-compositionality and non-determinism in transition rules}
    \label{fig:circular-rules}
\end{figure}
This rule is not compositional: the syntactic constructs in the premise are not immediate constituents of the construct in the conclusion. While non-compositionality is useful, if not necessary, in some cases, e.g. loop constructs, it can still give rise to unexpected scenarios, one particular case being rules with overlapping patterns.

From the standpoint of the definition of a specification, rule application is non-deterministic for rules with overlapping patterns, and the evaluation satisfies the semantics if at least one derived tree is closed. As an example, consider the formation rules of $e$ also include the terminal \sco{stop}, the semantics of which are given by rule STOP in Figure \ref{fig:circular-rules}. Evaluating the \sco{stop} construct is therefore non-deterministically defined as either diverging via rule CIRCULAR or eventually converging to 0, via rule STOP. However, when transcribing the specification as-is and evaluating \sco{stop}, the metalanguage interpreter will always diverge, owed to matching CIRCULAR first, and fail to consider rule STOP, terminating the evaluation.

\section{Conclusions and future work}
\label{sec:future-work}

We have presented a metalanguage and its interpreter that provide a lightweight method for expressing natural semantics specifications of small programming languages of the kind usually designed by undergraduates in their projects. Our central observation is that learners of natural semantics very often make mistakes that are type errors wrt. the notation used. Notably, the uniformity of transitions appears to be a threshold concept for learners.

Our metalanguage is therefore equipped with a type system that captures such errors. Moreover, the syntax is kept reasonably similar to textbook notation preventing students from having to learn separate notations. The type checker helps weed out inconsistencies in transition rules. Although experimental and not publication-quality, generated \LaTeX code can ease the process of writing reports, as an added benefit.

Regardless, there is room for many improvements e.g., investigating the viability of a polymorphic type system without affecting other features of the language e. g., partial function support. Evaluations and rule definitions could also benefit from improvements, notably in terms of warnings. As outlined in section \ref{simulate-interp}, rules may be non-exhaustive. Warning the user of such cases, similarly to  typed functional languages such as Haskell, would increase usability. Additionally, a termination checker similar to the one used in languages such as Agda \cite{Norell2009} could be employed to give warnings about non-compositional rules.

Other useful features include support for code generation of interpreters targeting other implementation languages as back-ends. The current iteration supports experimental generation of Haskell code, but not all use cases are properly handled yet, e. g. overlapping patterns in transition rule conclusions. For undergraduate students, a better choice for the back-end would be Java, to better integrate with tools like ANTLR\cite{Parr13} or SableCC\cite{DBLP:conf/tools/GagnonH98}.

An important next step is to introduce the tool into the setting of future teaching activities. Concretely, we plan to use the tool in forthcoming courses on programming language design and semantics. This will involve devising teaching activities that introduce students to the metalanguage. It is our opinion that such activities should emphasise the relationship between designing specifications and general purpose programming in order to facilitate the understanding of threshold concepts such as the uniformity of transitions.

Another topic for further work is to consider small-step semantics and the challenges that students face when learning this style of operational semantics. As for natural semantics, some of the problems with writing meaningful specifications can be seen as type safety violations.

\bibliographystyle{eptcs}
\bibliography{ITICSE}

\end{document}